\documentclass[fleqn, usenatbib, useAMS]{mnras}
\usepackage[fleqn]{amsmath}
\usepackage{amssymb}
\usepackage[none]{hyphenat}
\usepackage{times}
\usepackage{graphicx}
\usepackage{color}
\usepackage{blindtext}

\newcommand{\ias}{{\sc \tt IAS15}\xspace}
\newcommand{\reb}{{\sc \tt REBOUND}\xspace}
\newcommand{\rebx}{{\sc \tt REBOUNDx}\xspace}
\definecolor{amber}{rgb}{1.0, 0.49, 0.0}
\title[Ejection of rocky and icy material from binary star systems]{Ejection of rocky and icy material from binary star systems: Implications for the origin and composition of 1I/`Oumuamua}
\author[A.P. Jackson et al.]{Alan P. Jackson$^{1,2}$\thanks{E-mail: ajackson@cita.utoronto.ca}, Daniel Tamayo$^{1,3}$, Noah Hammond$^{1}$, Mohamad Ali-Dib$^{1,3}$, Hanno Rein$^{1}$\\%
$^1$Centre for Planetary Sciences, University of Toronto at Scarborough, 1265 Military Trail, Toronto, Ontario, M1C 1A4, Canada\\%
$^2$School of Earth and Space Exploration, Arizona State University, 781 E. Terrace Mall, Tempe, Arizona 85287, USA\\
$^3$Canadian Institute for Theoretical Astrophysics, 60 St. George St, University of Toronto, Toronto, ON M5S 3H8, Canada}
\date{Submitted 2017}
\pagerange{\pageref{firstpage}--\pageref{lastpage}}
\pubyear{2017}
\begin{document}
 \label{firstpage}
  \maketitle
 
 \begin{abstract}
  In single star systems like our own Solar system, comets dominate the mass budget of bodies that are ejected into interstellar space, since they form further away and are less tightly bound.  However 1I/`Oumuamua, the first interstellar object detected, appears asteroidal in its spectra and in its lack of detectable activity.  We argue that the galactic budget of interstellar objects like 1I/`Oumuamua should be dominated by planetesimal material ejected during planet formation in circumbinary systems, rather than in single star systems or widely separated binaries. We further show that in circumbinary systems, rocky bodies should be ejected in comparable numbers to icy ones. This suggests that a substantial fraction of additional interstellar objects discovered in the future should display an active coma.  We find that the rocky population, of which 1I/`Oumuamua seems to be a member, should be predominantly sourced from A-type and late B-star binaries.
 \end{abstract}


 \section{Introduction}
 \label{sec:intro}
 
 With the discovery of 1I/2017 U1 (`Oumuamua) we now have our first glimpse of an interstellar object \citep{meech2017}.  With a velocity at infinity of around 26~km/s, and an inclination of 122$^{\circ}$ that precludes a close encounter with one of the Solar system planets\footnote{Orbital parameters taken from the JPL small body database, \url{ssd.jpl.nasa.gov/sbdb.cgi}}, 1I/`Oumuamua is securely of interstellar origin.  Furthermore, \citet{mamajek2017} showed that the trajectory of 1I/`Oumuamua prior to encountering the Solar system is not consistent with a recent ejection from a nearby star and that its velocity relative to the galactic background is close to the local standard of rest.  This suggests that 1I/`Oumuamua was ejected at low speed from its parent system and that it has been wandering interstellar space for a long time since.
 
 The existence of 1I/`Oumuamua can also be used to place constraints on the mass of material typically ejected by planetary systems, \citep[e.g.][]{laughlin2017, raymond2017}, albeit that with only a single object such estimates are subject to large uncertainties.  In addition while the orbital characteristics of 1I/`Oumuamua are consistent with our expectations for an interstellar object, its physical characteristics are rather more surprising, in particular its lack of observable activity \citep[e.g.][]{jewitt2017} and highly elongated shape \citep[e.g.][]{bolin2017, meech2017}.
 
 \citet{cuk2017} recently noted that binary and multiple star systems could be a major source of interstellar bodies and suggested that 1I/`Oumuamua might have originated as a fragment of a much larger body that was tidally disrupted during ejection from its parent system.
 
 In this work we quantitatively examine this scenario, showing that tidal disruptions are unlikely, but that tight binary systems can nevertheless eject an amount of rocky material comparable to the predominantly icy material thrown out by single and wide binary star systems.  We begin in Section~\ref{sec:comp} by summarising the current state of knowledge of the composition of 1I/`Oumuamua, in particular examining whether it is a rocky or icy object since this has important implications for its origin.  In Section~\ref{sec:ejection} we outline our rationale for expecting that binary systems may be a major source of ejected material and then describe our methods in Section~\ref{sec:method}.  We present the results of our analysis in Section~\ref{sec:results} and discuss their implications, before summarising in Section~\ref{sec:conc}.
 
 \section{The composition and nature of 1I/`Oumuamua}
 \label{sec:comp}
 
 Three clues to the composition of 1I/`Oumuamua are its spectral reflectance profile, lack of a coma and elongated shape.
 A number of groups have obtained photometric colours across a variety of bands in the visible and near infrared \citep[e.g.][]{bannister2017, bolin2017, jewitt2017, masiero2017, meech2017, ye2017}.  These observations reveal an object with a relatively constant shallow red slope to its reflectance in the visible and near infrared.  It is redder than some inner Solar system asteroid classes and does not appear to demonstrate the broad absorption longwards of 0.75~$\mu$m seen in many asteroid classes.  However it is also significantly less red than many Kuiper belt objects.  Rather it seems fairly similar to D-type asteroids or the nuclei of long period comets.
 
 Despite its spectral similarity to volatile rich objects,
 1I/`Oumuamua has shown no detectable level of activity.  \citet{jewitt2017} place an upper limit on the rate of mass loss due to activity at $\sim 2\times 10^{-4}$~kg~s$^{-1}$, limiting the area of exposed water ice on the surface to $<$1~m$^2$.  As they point out however, this does not preclude the existence of water ice within the interior of 1I/`Oumuamua since regolith can be an extremely good insulator (with thermal diffusivities as low as $\sim10^{-8}$~m$^2$~s$^{-1}$) and as such the thermal impulse during its transit of the inner Solar system may only have penetrated around 0.5-1~m below the surface at the present time. 
 
 One might propose that since 1I/`Oumuamua has spent a very long time outside the envelope of a stellar magnetosphere, exposed to the background galactic radiation environment, it might have developed a thick volatile depleted layer \citep[e.g.][]{fitzsimmons2018}. We find this unlikely however, as the nuclei of long period comets spend the majority of their time in the outer reaches of the Oort cloud, experiencing the same interstellar radiation field that 1I/`Oumuamua would have done, yet they become active within the inner solar system, an issue also acknowledged by \citet{fitzsimmons2018}. Additionally, detailed calculations have shown that interstellar radiation is only capable of altering the upper few centimetres, even after over 10$^9$~years beyond the Solar heliopause, where exposure is highest to the most chemically significant radiation \citep{cooper2003}. This penetration depth would be sufficient to change the colour of the surface, but probably not sufficient to generate an insulating layer that could keep volatiles from degassing during passage through the inner Solar system.
 
 As such we conclude that the lack of activity from 1I/`Oumuamua is unlikely to be related to radiation exposure during its long sojourn in interstellar space.  Instead we argue that the lack of activity is related to its origin: either it is a truly volatile poor body, or it underwent sufficient processing within its parent system to generate a thick ($\ga$0.5~m) insulating crust.  Both of these options imply that 1I/`Oumuamua spent a significant period of time within the inner parts of its parent system prior to being ejected.
 
 \section{The case for binary systems as a major source of ejected material}
 \label{sec:ejection}
 
 If we consider a star with a companion and a population of small bodies, simple analysis shows that for ejection to dominate over accretion in encounters between the small bodies and the companion, the escape velocity of the companion should exceed the Keplerian orbital velocity at its semi-major axis \citep[e.g.][]{wyatt2017}.  This is independent of whether the companion is a planet or a star.  In the Solar system Jupiter, Saturn, Uranus and Neptune all satisfy this criterion, however as discussed by \citet{wyatt2017} the timescale for ejection is also important.  For Neptune and Uranus, while ejection dominates over accretion, bodies can take 100s of millions of years to be ejected, long enough for galactic tides to perturb the bodies such that they enter the Oort cloud.  Moreover, as we discussed in Section~\ref{sec:comp} 1I/`Oumuamua appears to be rocky or devolatilised, suggesting it was ejected from inside the ice line.
 
 For a solar-mass star, efficient ejection inside typical ice line distances of a few AU within the first few Myr of the life of the system requires that the companion has a mass greater than that of Saturn.  However, radial velocity surveys show that the occurrence rate of giant planets at orbital periods of 100-400 days is low---approximately 3\% \citep{santerne2016} rising to around 10\% for orbital periods of $<$10 years \citep{mayor2011}.  As such, we expect that at most 10\% of Sun-like single stars will host a planet capable of efficiently ejecting material interior to the ice line.  \citet{laughlin2017} and \cite{raymond2017} thus argue that if 1I/`Oumuamua is indeed rocky, then typical extrasolar asteroid belts must be unusually massive.\footnote{While material can still be ejected from our own inner Solar system as a result of the terrestrial planets passing material out to Jupiter, the timescales for doing so are longer ($\sim$10-100~Myr), which means that the total mass can be significantly depleted by collisional processing prior to ejection \citep{jackson2012, shannon2015}.}  Similarly, recent results from micro-lensing surveys \citep[e.g.][]{suzuki2016, mroz2017} suggest that giant planets at larger separations are also not common.
 
 While giant planets are relatively uncommon, tight binary systems are abundant \citep{duchene2013}, and are extremely efficient at ejecting material \citep{smullen2016}. They may therefore represent a dominant source of interstellar small bodies. We now examine this hypothesis in detail.
 
 \section{Method}
 \label{sec:method}
 
 We want to be able to examine the ejection of small bodies from binary systems, and compare this with single stars, in a way that is consistent across spectral classes. As summarized by \citet{duchene2013}, the properties of binary systems, in particular their separation and their multiplicity frequency, are dependent on the mass of the primary, though they note that these dependences are subject to a fair degree of uncertainty and are the subject of active investigation.  As such, we need to take into account the relative abundance of stars of different masses to be able to build a complete picture of the binary population. To do this we construct a population synthesis model as detailed below.
 
 Physically, our picture is one of planetesimals migrating inwards during the early phases of planet formation, in the presence of a protoplanetary disk.  \cite{holman1999} showed that any material in circumbinary orbit migrating inward will become unstable on short timescales once it passes a stability boundary $a_{\rm c, out}$, for which they provide an empirical fit to results from N-body simulations (their equation 3).  This critical distance is a function of the binary mass ratio and eccentricity and ranges from around 2 to 4 times the binary separation.  We thus envisage planetesimals migrating in and then being ejected once they pass $a_{\rm c, out}$.
 
  \begin{table*}
     \centering
     \caption{Binary fractions and separation distributions used in the population synthesis model.  Chosen to match \citet{duchene2013}}
     \label{table:bindists}
     \begin{tabular}{r|c|l}
        Mass ($M_{\odot}$) & Binary fraction (\%) & Separation distribution \\
         $0.1 \leq M_1 < 0.6$ & 26 & log-normal, mean = 5.30~AU, $\sigma_{\log_{10} a}$ = 0.867\\
         $0.6 \leq M_1 < 1.4$ & 44 & log-normal, mean = 44.7~AU, $\sigma_{\log_{10} a}$ = 1.53\\
         $1.4 \leq M_1 < 6.5$ & 50 & log-normal, mean = 0.141~AU, $\sigma_{\log_{10} a}$ = 0.50, + log-normal, mean = 316~AU, $\sigma_{\log_{10} a}$ = 1.65\\
         $M_1 \geq 6.5$       & 70 & log-normal, mean = 0.178~AU,  $\sigma_{\log_{10} a}$ = 0.36, + log-uniform, min = 0.50~AU, max = 10$^6$
     \end{tabular}
 \end{table*}

 \subsection{Population synthesis model}
 \label{sec:method:popsynth}
 
 Our population synthesis model proceeds as follows for the construction of a single system:
 
 \begin{enumerate}
     \item Sample the system mass, $M_{\rm sys}$, from the system initial mass function of \citet{chabrier2003}.
     \item Sample the mass ratio, $\mu = M_2/(M_1+M_2)$ uniformly in the range [0, 0.5], as suggested by \citet{duchene2013}, and use this to calculate $M_1$ and $M_2$.
     \item Determine which of the \citet{duchene2013} mass bins  $M_1$ falls into and use the appropriate multiplicity fraction to determine if the system is actually binary\footnote{Hierarchical triples would act similarly to binaries in our models, so for simplicity we ignore multiple systems.}. If the system is not binary set $\mu=0$, $M_1=M_{\rm sys}$ and return to step (i) and begin again for the next system.
     \item Sample the eccentricity uniformly in the range [0, 0.9]\footnote{We cap the eccentricity at 0.9 to avoid computationally expensive N-body simulations.} and the binary separation from the distributions listed in Table~\ref{table:bindists}, taken to approximately match \citet{duchene2013}.
     \item Determine the radii and luminosities of the stars by interpolating on the isochrones generated by MIST (the MESA Isochrones and Stellar Tracks, \citealt{dotter2016, choi2016}) for a system age of 1~Myr.\footnote{We take 1~Myr to be a representative age of a young system in which substantial disk migration may be ongoing.}
 \end{enumerate}
 
 Our population synthesis model ensures that binary systems of different masses have the correct weighting and also simultaneously constructs a single star comparison population.  Integrated over all stellar masses the model produces binary systems at a rate of 30\% by number, while the higher multiplicity rates for more massive stars means that binaries constitute 41\% of all mass.
 
 We assume that disk dynamics are broadly the same across all systems and that disk masses are a roughly constant fraction of the total stellar mass, such that the mass of material that is ejected by each binary is a constant fraction of the system mass.  As such we will weight systems by their total mass for all of our analysis in the later sections.
 
 In examining our binary systems we should also take into account that binaries with very wide separations are unlikely to have substantial material in circumbinary orbit, but rather can be expected to behave like single stars.  Exactly where to draw the boundary between a wide binary in which we expect the stars to have their own disks that are largely independent and close binaries in which we expect a dominant circumbinary disk is not clear.  We choose to set the boundary as the point at which the outermost stable orbit around the larger star, $a_{\rm c, in}$ (as determined by Eq.~1 of \citealt{holman1999}), is greater than 10~AU.  Our results are not sensitive to the precise choice of this cut-off however, as discussed in Section~\ref{sec:results}. Given these assumptions, we estimate a mass fraction of 26\% in circumbinaries, with the remainder in singles and wide binaries.
 
 One could envisage there being some intermediate separation binaries in which a circumprimary disk extends to $a_{\rm c, in}$ such that material in the outer regions of the disk can be ejected due to viscous spreading across the $a_{\rm c, in}$.  The mass available for ejection in such a system is fairly limited however, being sourced from the lower density outer regions of the disk, whereas inward drift can potentially make a large fraction of the mass of a circumbinary disk available for ejection.  As such we focus here on circumbinary systems.
 
 \subsection{$N$-body simulations}
 \label{sec:method:nbody}
 
 While an object that migrates inward past $a_{\rm c, out}$ will become unstable on short timescales, this does not determine the fate of the object, which can either be ejected or accreted onto one of the stars.  In addition we also want to examine the distribution of close encounter distances prior to ejection to assess the possible role of tidal disruptions as suggested by \citet{cuk2017}.  Since the unstable region inside $a_{\rm c, out}$ also applies to gas, we expect that the disk will have a central cavity where the gas density is low. The fate of bodies migrating into this central unstable region can thus be determined using simple $N$-body integrations.
 
 We conduct a set of 2000 $N$-body simulations using the high-order adaptive-timestep integrator \ias in the \reb integration package \citep{ReinSpiegel2015,ReinLiu2012}.  As described in Section~\ref{sec:method:popsynth} the mass ratio is uniformly distributed on~[0,~0.5] and the eccentricity is uniformly distributed on~[0,~0.9]. We then initialize particles on co-planar and initially circular orbits at orbital distances beyond the instability limit from \citet{holman1999}. Finally, we apply an inward drag force that (in an orbit-averaged sense) exponentially decreases the semimajor axis on a timescale 1000 times longer than the orbital period using the \texttt{modify\_orbits\_forces} routine in \rebx, and integrate for $10^4$ binary orbits. We track ejections by flagging particles that go beyond 100 binary orbit separations.
 
 Since Newtonian gravity is scale invariant, we express masses in terms of the primary mass and distances in units of the binary separation. Collisions introduce the scale of the stellar radii into the problem, so we first run a universal set of 2000 integrations of point particles, and dimensionalize after the fact, drawing stellar masses, radii, and binary separations as described above.  We then look for collisions with the stars in this post-processing step.
 
 \section{Results and discussion}
 \label{sec:results}
 
 Comparing the distribution of collisional and ejection outcomes for our $N$-body simulations we find that the fraction of particles ejected is near unity, reaching a minimum of $95\%$ for O and B stars that have the largest collisional cross-sections. As such, it is a good approximation to assume that all material that migrates to within the critical semi-major axis will be ejected.
 
 We now generate a synthetic population of 10$^7$ systems using our population synthesis models.  As noted in Section~\ref{sec:method:popsynth} we assume that the mass of material that is ejected by each binary is a constant fraction of the system mass.  For now we remain agnostic as to what this ejection fraction is and label it as $M_{\rm ej}^{\rm bin}$.  We discuss possible values for $M_{\rm ej}^{\rm bin}$ and their implications in Section~\ref{sec:results:interstellarpop}.
 
 As we discussed above, we believe that 1I/`Oumuamua is either rocky or has a substantial devolatilised crust, implying that it spent a considerable time inside the ice line in its parent system.  Since bodies are rapidly ejected once they move within $a_{\rm c, out}$ we are interested specifically in the subset of circumbinary systems where the ice line is outside $a_{\rm c, out}$.
 
 We determine the ice line distance, $a_{\rm ice}$, as the distance at which the radiative equilibrium temperature of a blackbody is 150~K, assuming the total luminosity of the two stars is located at the centre of mass of the system.  While the presence of a massive protoplanetary disk can potentially have a large effect on the ice line location through optical depth preventing stellar radiation reaching the interior and through viscous heating, we note that the region interior to $a_{\rm c, out}$ is expected to have a low gas density such that these effects can be ignored in that region.  As such, while the precise location of the ice line might vary if it is outside $a_{\rm c, out}$ the fraction of systems with $a_{\rm ice} > a_{\rm c, out}$ determined in this way should provide a reasonable estimate.
 
 Since we are allowing $a_{\rm ice}$ to be close to $a_{\rm c, out}$ it is worth considering how long a body might need to be inside $a_{\rm ice}$ to develop a substantial devolatilised layer.  For Solar system comets mass loss rates are typically a few $10^{-7}$~kg~m$^{-2}$~s$^{-1}$ just inside the ice line \citep{ahearn2012}, corresponding to an erosion rate of $\sim$0.01~m~yr$^{-1}$.  These mass loss rates rise very rapidly as orbital distance decreases however, reaching $\sim10^{-5}$-$10^{-4}$~kg~m$^{-2}$~s$^{-1}$ (1-10~m~yr$^{-1}$) by 1~AU where Solar intensity is twice as high as at the ice line.  As such we might expect to require a few centuries to perhaps a few millenia of continuous activity to generate a sufficiently thick devolatilised crust, which is very plausible for a 100~m cometary body migrating in via gas drag from close to $a_{\rm c, out}$.
 
  \begin{figure}
     \centering
     \includegraphics[width=\columnwidth]{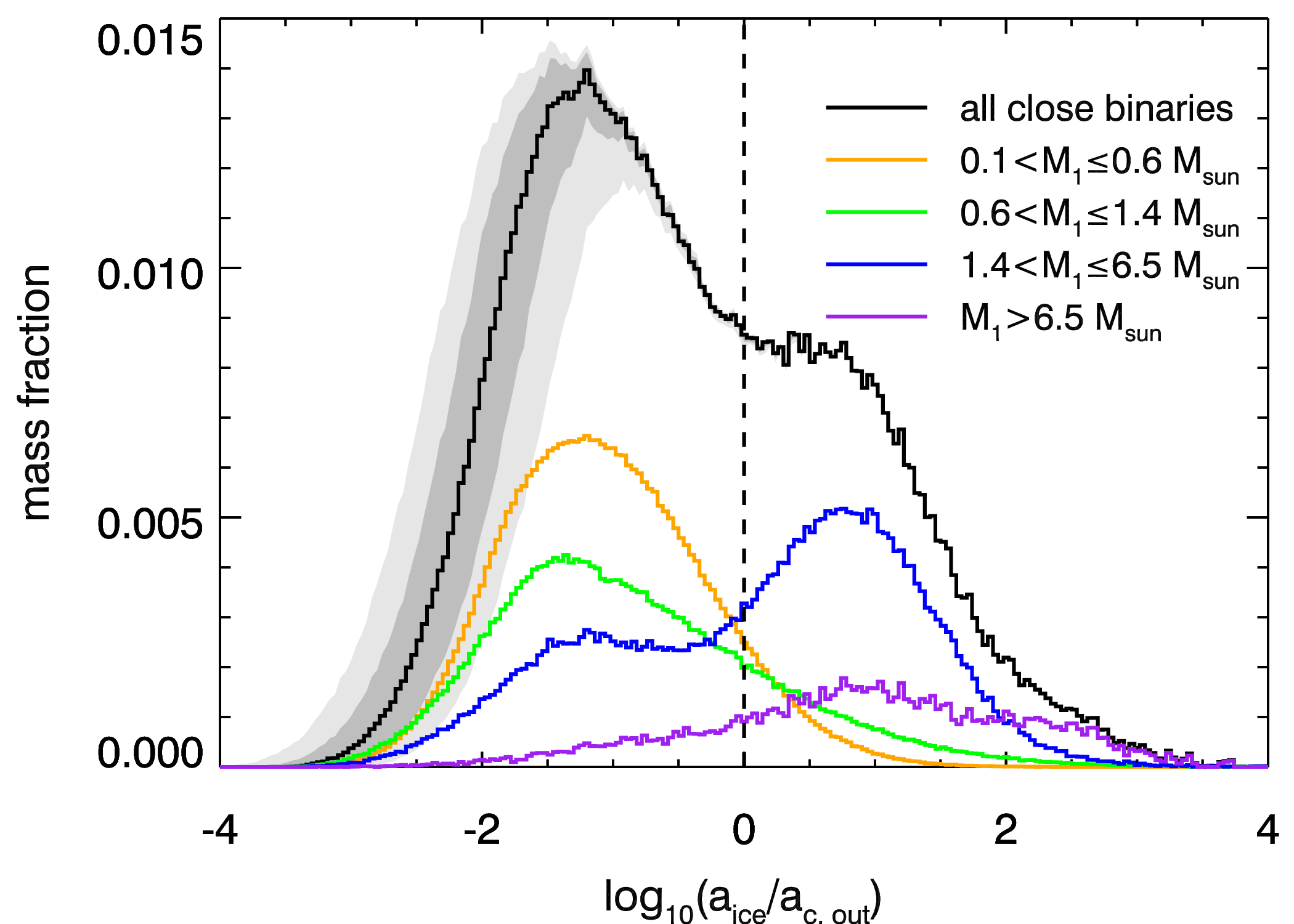}
     \caption{Histogram of $a_{\rm ice}/a_{\rm c, out}$ for all close binaries, weighted by the mass of the system.  The total distribution is shown in black, while the orange, green, blue and purple curves show the contributions from stars of different masses.  The grey shading shows the effect of changing the wide binary cut-off to $a_{\rm c, in} >$ 2.5, 5, 20, or 40~AU.  All curves are normalised to the total mass of close binaries for a wide binary cut-off of $a_{\rm c, in} >$ 10~AU.  The dashed line indicates $a_{\rm ice}/a_{\rm c, out}=1$.}
     \label{fig:aiac}
 \end{figure}
 
 In Fig.~\ref{fig:aiac} we then show the distribution of $a_{\rm ice}/a_{\rm c, out}$, weighted by the mass of the system.  With our assumption that the mass of material ejected relative to the system mass is the same for all binaries the fraction with $a_{\rm ice}/a_{\rm c, out} > 1$ in this distribution then tells us the fraction of interstellar material ejected by binaries that will be rocky or devolatilised.  We find that this fraction is 36\%, such that the ratio of icy to rocky/devolatilised material is roughly 2:1.  We also show the contributions from stars of different masses in Fig.~\ref{fig:aiac}.  This shows that the population of icy interstellar material predominantly originates from low mass stars while the population of rocky/devolatilised material is dominated by intermediate mass stars.  The fraction of systems with $a_{\rm ice}/a_{\rm c, out} > 1$ is relatively insensitive to the choice we made in Section~\ref{sec:method:popsynth} regarding where to place the wide binary cut-off, as shown by the grey shading in Fig.~\ref{fig:aiac}.
 
 \subsection{The population of interstellar bodies}
 \label{sec:results:interstellarpop}
 
 The total population of interstellar bodies will be the combination of those ejected by binary systems and those ejected by single star systems.  We previously defined the fractional mass ejected by binary systems as $M_{\rm ej}^{\rm bin}$.  For single star systems (and wide binaries) we assume that ejection of significant masses of material is limited to those stars that host giant planets, and that those systems eject a fraction of their mass equal to $M_{\rm ej}^{\rm sin, gp}$.  Averaged over all singles the fractional mass ejected is then $f_{\rm giant}M_{\rm ej}^{\rm sin, gp}$, where $f_{\rm giant}$ is the fraction of singles that host giant planets.  The total fractional mass averaged over all stars is then
 \begin{equation}
     M_{\rm interstellar} = f_{\rm bin} M_{\rm ej}^{\rm bin} + (1 - f_{\rm bin}) f_{\rm giant}M_{\rm ej}^{\rm sin, gp},
     \label{eq:minter1}
 \end{equation}
 where $f_{\rm bin}$ is the mass fraction of stars that are binaries.  In Section~\ref{sec:method:popsynth} we found that $f_{\rm bin}$ = 26\%, while in Section~\ref{sec:ejection} we argued that $f_{\rm giant}$ is no more than 10\%.  Using these values Equation~\ref{eq:minter1} becomes
  \begin{equation}
     M_{\rm interstellar} = 0.26 M_{\rm ej}^{\rm bin} + 0.074 M_{\rm ej}^{\rm sin, gp}.
     \label{eq:minter2}
 \end{equation}
 We are also interested in the fraction of all interstellar bodies that are rocky or devolatilised rather than icy, $R_{\rm interstellar}$.  We thus divide the ejected masses into rocky and icy components such that
\begin{equation}
   \label{eq:minter3}
   \begin{split}
        R_{\rm interstellar} & = \frac{M_{\rm interstellar, rock}}{M_{\rm interstellar}} \\
            & = \frac{0.26 R_{\rm bin}  M_{\rm ej}^{\rm bin} + 0.074 R_{\rm sin} M_{\rm ej}^{\rm sin, gp}}{0.26 M_{\rm ej}^{\rm bin} + 0.074 M_{\rm ej}^{\rm sin, gp}},
   \end{split}
 \end{equation}
 where $R_{\rm bin}$ and $R_{\rm sin}$ are the rock/ice fractions for binaries and single systems respectively.
 
 The Nice model for the early evolution of the outer Solar system suggests that the Solar system began with an outer planetesimal disk of $\sim$30~$M_{\oplus}$, the majority of which was ejected \citep[e.g.][]{gomes2005, levison2008}.  In addition perhaps 1~$M_{\oplus}$ was ejected from the inner Solar system \citep[e.g.][]{shannon2015}.  If other systems that host giant planets behave in a similar way to the Solar system then, taking a median value we can expect that $M_{\rm ej}^{\rm sin, gp} \sim 30 M_{\oplus}/M_{\odot}$ and that $R_{\rm sin} \sim 0.033$.
 
 We must now estimate $M_{\rm ej}^{\rm bin}$.  In doing so we first note that from our $N$-body simulations we expect binaries to eject essentially all material that migrates to within $a_{\rm c, out}$.  Moreover since the ejection mechanism only relies on the central binary itself, material can be ejected from very early in the life of the system, whereas in a single system ejection can only begin once a giant planet has formed.  If most systems begin with disks close to the gravitational instability limit of $\sim$0.1~$M_{\rm sys}$ and a typical gas to dust ratio of around 100 this implies a total mass of solids of $\sim$300~$M_{\oplus}/M_{\odot}$.  If we estimate that around 10\% of this material migrates to within the critical stability radius this implies that every binary system ejects as much material as a single star system that hosts giant planets.  Using $M_{\rm ej}^{\rm bin} \sim 30 M_{\oplus}/M_{\odot}$ as our fiducial value we then find that $M_{\rm interstellar} \sim 10 M_{\oplus}/M_{\odot}$.  Since we expect $R_{\rm bin}$ = 0.36 this leads to $R_{\rm interstellar} \sim 0.29$.  With this estimate more than three quarters of interstellar bodies originate from binary stars, while for rocky objects the fraction would be even higher.
 
 \subsection{Tidal disruptions vs extreme heating}
 \label{sec:results:disruption}
 
 \citet{cuk2017} suggested that 1I/`Oumuamua might have originated as a tidal disruption event of a planet. We can test this hypothesis with our $N$-body simulations. For massive stars, the Roche limit (inside which tidal disruptions will occur) is smaller than the stellar radius. To provide the best-case scenario for tidal disruptions, we re-calculate the stellar collision outcomes using stellar radii for the Zero Age Main Sequence, when the radii are at a minimum.  Treating a gravity-dominated planet as a strengthless body, we compute the Roche radius as $R_{\rm Roche} = 1.26 (\rho_*/\rho_{p})^{(1/3)} R_*$, where we assume a lower bound for the density of the planet of $\rho_{p}$ = 3000~kg~m$^{-3}$. We find that none of our 2000 simulations results in a close encounter within the Roche radius.  Indeed the majority of closest approach distances are far outside the Roche radius.  This implies that the frequency of tidal disruptions is $\lesssim 10^{-3}$ times the ejection rate. Note that there is no contribution from higher mass stars since they always have their Roche radius inside the star.
 
 \begin{figure}
     \centering
     \includegraphics[width=\columnwidth]{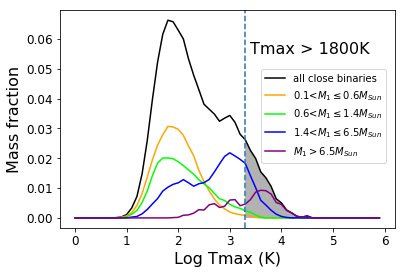}
     \caption{Mass weighted distribution of peak temperatures reached by particles ejected in our $N$-body simulations.}
     \label{fig:peaktemps}
 \end{figure}
 
 While we have shown that tidal disruptions are rare,  material that is ejected can potentially be heated to high temperatures during a close approach.  In Fig.~\ref{fig:peaktemps} we show the mass weighted distribution of peak temperatures experienced by particles in our $N$-body simulations before they are ejected.  Around 10\% of bodies experience peak temperatures in excess of 1800~K, sufficient to melt rock.  These are drawn from systems independent of their ice line distances, though we note that systems with higher temperatures at the stability boundary $a_{\rm c, out}$ are more likely to experience extreme heating. 
 It would be interesting to consider in future work how extreme heating may modify the shape and surface layers of 1I/`Oumuamua.  We do note that the extreme elongation would be easier to maintain for an annealed, monolithic body than for a rubble pile.  The fact that such heating predominantly occurs around high-mass stars also further motivates understanding the planet formation process in these extreme regimes.

 Since submission of this work, the revised manuscript of \citet{cuk2017} points out that planets on circumprimary (S type), rather than the circumbinary (P type) orbits considered here may be tidally disrupted more effectively. As he points out, a dynamical instability of several planets around a single star would be required to scatter the doomed planet into a region of phase space where it can chaotically transfer between stars and suffer a close enough encounter for tidal disruption. It would be valuable to extend the analysis of \citet{cuk2017} to quantify the fraction of instabilities that could plausibly deliver planets into this chaotic intermediate regime instead of directly ejecting them. 

 Alternatively, planets in such an S type configuration could have their eccentricities secularly driven to high enough values by the binary companion \citep{naoz2012, petrovich2015} for tidal disruption, though this slow secular driving should be quenched by the tidal forces themselves before the planet is able to come close enough to the star for breakup \citep[e.g.][]{wu2011, liu2014}.
 
 \section{Conclusions}
 \label{sec:conc}
 
 We have shown that the population of interstellar objects can be dominated by planetesimals ejected during planet formation in circumbinary systems.  Even if a typical circumbinary only ejects as much material as the Solar system we would still expect close binaries to be the source of more than three quarters of interstellar bodies due to the relatively low abundance of single star systems with giant planets like the Solar system.  Whereas in the Solar system the ejected material is overwhelmingly icy, we expect that around 36\% of binaries may predominantly eject material that is rocky or substantially devolatilised, leading to similar expectations for the abundance of rocky/devolatilised bodies in the interstellar population.
 
 Within the close binary population the dominant source of rocky/devolatilised material are intermediate mass stars, A-stars and late B-stars.  As such, we suggest that the apparently rocky or devolatilised appearance of 1I/`Oumuamua indicates that it likely originated in such an intermediate mass binary system.
 
 We find that tidal disruptions during ejection from binary systems, as suggested by \citet{cuk2017} are rare.  While close encounter distances that result in strong tidal effects are highly unusual around 10\% of bodies may experience extremely high peak temperatures during ejection.  How these extreme temperatures could influence the shape and surface layers of a body like 1I/`Oumuamua would be an interesting topic for future work, and that such heating predominantly occurs around high-mass stars also further motivates understanding the planet formation process in these extreme regimes.

 \section*{Acknowledgements}

APJ thanks Andrew Shannon and Casey Lisse for helpful discussions.
APJ gratefully acknowledges funding through NASA grant NNX16AI31G (Stop hitting yourself). 
DT thanks Greg Laughlin for insightful discussions.
HR gratefully acknowledges funding through NSERC Discovery Grant RGPIN-2014-04553.
The authors thank the referee for a rapid review and helpful comments that have improved the manuscript.

{\footnotesize
\bibliographystyle{mnras}
\bibliography{interstellarasteroids}
}
 
 \label{lastpage}
\end{document}